\documentstyle[twoside,12pt]{article}
\newcommand{\nc}{\newcommand}
\nc{\la}{\lambda} \nc{\La}{\Lambda}
\nc{\al}{\alpha}
\nc{\te}{\theta}  \nc{\be}{\beta}
\nc{\ga}{\gamma}  \nc{\Ga}{\Gamma}
\nc{\de}{\delta}  \nc{\De}{\Delta}
\nc{\si}{\sigma}  \nc{\ka}{\kappa}
\nc{\om}{\omega}  \nc{\Om}{\Omega}
\nc{\nf}{\infty}   \nc{\nl}{\newline}
\nc{\ra}{\longrightarrow}
\nc{\beq}{\begin{equation}}
\nc{\eeq}{\end{equation}}
\nc{\beqa}{\begin{eqnarray}}
\nc{\dst}{\displaystyle}
\nc{\eeqa}{\end{eqnarray}} \nc{\nnb}{\nonumber}
\title{\bf Compact Einstein-Weyl
\\ four-dimensional manifolds}
\author{Guy Bonneau\thanks {\noindent Laboratoire de Physique
Th\'eorique et des Hautes Energies,
 Unit\'e associ\'ee au CNRS UMR 7589, Universit\'e Paris 7,
 2 Place Jussieu, 75251 Paris Cedex 05. bonneau@lpthe.jussieu.fr}}
\begin{document}
\maketitle
\begin{abstract}
\noindent We look for
 four dimensional Einstein-Weyl spaces equipped with a regular
Bianchi metric. Using the explicit 4-parameters expression of the
distance obtained in a previous work for non-conformally-Einstein
Einstein-Weyl structures, we show that only four 1-parameter
families of regular metrics exist on orientable manifolds : they
are all of Bianchi
$IX$ type and conformally K\"ahler ; moreover, in agreement with
general results, they have a positive  definite conformal scalar
curvature. In a Gauduchon's gauge, they are compact and we obtain
their topological invariants. Finally, we compare our results to
the general analyses of Madsen, Pedersen, Poon and Swann : our
simpler parametrisation allows us to correct some of their
assertions.
\end{abstract}
$$ $$ PACS : 02.40.Hw, 04.20.Jb


\vfill {\bf PAR/LPTHE/98-25/gr-qc/9806037}\hfill  Revised version
: October 1998
\newpage

\section{Introduction} In the last years, Einstein-Weyl geometry
has deserved some interest. A Weyl space \cite{PS93} is  a
conformal manifold with a torsion-free connection D and a one-form
$\ga$ such that for each representative metric g in a conformal
class [g],
\beq\label{a1}
 D_{\mu} g_{\nu \rho} = \ga_{\mu} g_{\nu \rho}\ .
\eeq  A different choice of representative metric : $g\ \ra\
\tilde{g} = e^f g$ is accompanied by a change in
$\ga\ : \ga\
\ra\ \tilde{\ga} = \ga + df\ .$  Conversely, if the one-form $\ga$
is exact, the metric g is conformally equivalent to a Riemannian
metric
$\tilde{g}$ : $ D_{\mu}\tilde{ g}_{\nu \rho} = 0.$ In that case,
we shall speak of an {\it exact} Weyl structure. Einstein-Weyl
spaces are defined by :
\beqa\label{a5}  {\cal R}^{(D)}_{(\mu\nu)}  & = &
\La ' g_{\mu\nu}\
\ \Leftrightarrow \nnb \\
 R^{(\nabla)}_{\mu\nu} +  \nabla_{(\mu}\ga_{\nu)} +
\frac{1}{2}\ga_{\mu}\ga_{\nu} & = & \La\  g_{\mu\nu}\ \ ,\ \ \La  =
\La' - \frac{1}{2}[\nabla_{\la}\ga^{\la} -
\ga_{\la}\ga^{\la}] \ .
\eeqa  This condition is a (conformally invariant) generalisation
of the usual Einstein condition. Note that for an exact
Einstein-Weyl structure, $\ga = df$, the representative metric is
conformally Einstein (for recent reviews see
refs.\cite{{PS93},{Calderbank},{Madsen-a},{mpps}}).   Let us also
recall that, {\bf in the compact case}, on general grounds, strong
results have been known for some time :
\begin{itemize}
\item There exits a unique metric in a given conformal class [g]
such that the Weyl form is co-closed \cite{Gauduchon}\ , $$
\nabla_{\mu}\ga^{\mu} = 0\ .$$ One then speaks of the
``Gauduchon's gauge" and of a ``Gauduchon's metric".
\item The analysis of Einstein-Weyl equations in this gauge gives
two essential results :
\begin{itemize}
\item  The dual of the Weyl form $\ga$ is a Killing vector
\cite{Tod92}:
$\ \nabla_{(\mu}\ga_{\nu)} = 0\ ,$
\item Four dimensional Einstein-Weyl space have a constant
conformal scalar curvature
\cite{{PedTod92},{Gauduchonbis},{PDSRheine}}.
\end{itemize}
\end{itemize}

\noindent On the other hand, in a recent work \cite{Bonneau97},
we studied in a purely local approach, (non-)compact 4-dimensional
Einstein-Weyl structures on cohomogenity-one  manifolds with a 3
dimensional group of isometries transitive on codimension-one
surfaces, {\it i.e.}, in the general relativity terminology,
Bianchi metrics
\cite{{DS95},{Tod95}}, and neither require compactness nor a
diagonal form for the metric. We however got interesting results
and proved that for {\it all class A Bianchi metrics}, there exits
a simple Gauduchon's gauge such that the conformal scalar
curvature is constant on the manifold. Our results are summarised
in a : \nl {\bf Theorem :} {\it (Non-)compact Einstein-Weyl
Bianchi metrics of the types $VI_0,\ VII_0,\ VIII$ and $IX$ are
conformally K\"ahler or conformally Einstein. In the non-exact
Einstein-Weyl case, the metric may be taken in a diagonal form and
is given in equ. (\ref{3-47}). The conformal scalar curvature has
a constant sign on the manifold and, in our particular Gauduchon's
gauge, the dual of the Weyl form is a Killing vector.}
\beqa\label{3-47} ds^2 & = & \frac{2}{\Ga}\left [\frac{k -n^{33}x
}{\Om^2(x)[1 + x^2]^2}(dx)^2
 + \frac{k - n^{33}x }{[1 + x^2]}[(\si^1)^2 + (\si^2)^2] +
\frac{\Om^2(x)}{k - n^{33}x }(\si^3)^2 \right] \ ,\nnb \\
\ga & = & \pm\frac{2\Om^2(x)}{k - n^{33}x }\si^3\
\ ;\ \ d\si^i =
\frac{1}{2}n^{ii}\epsilon_{ijk}\si^j\wedge\si^k\
\ \ ,\ \ \ i,j,k = 1,2,3\ \ ,\nnb \\ n^{11} & = & n^{22} = 1\ ,\
n^{33} = 0\ ({\rm Bianchi}\ VII_0),\ -1\ (VIII),\ +1\ (IX)\ \ ;
\nnb \\
\Om^2(x)  & = & n^{33} +
[n^{33}(x^2 - 1) -2 kx][l_1 + l_2\arctan(x)] + l_2[n^{33}x - 2
k]\,.
\eeqa Let us recall from  \cite{Bonneau97} that
 $\arctan(x)$ belongs to ($-\pi/2,\ +\pi/2$), thanks to the
integration constant
$l_1$; moreover,
$l_1,\ l_2,\ k$ and the homothetic positive parameter $\Ga$ are
arbitrary constants, and, in our particular Gauduchon's gauge, the
conformal scalar curvature is the constant
\beq\label{3-48} S^D = 4\Ga l_2\ .
\eeq

\noindent It is then natural to look for regular, complete metrics
among our solutions. The further requirement of compactness would
{\it a priori} be a gauge dependent condition : as we shall see,
all complete metrics in our Gauduchon's gauge are in fact compact
ones. This confirms that our choice does correspond to the unique
 Gauduchon's gauge of \cite{Gauduchon}.
\nl\noindent In this work, using the terminology of Gibbons and
Hawking
\cite{GH} on nuts and bolts, well adapted to the analysis of the
completeness of our candidate metrics on {\bf orientable}
manifolds, we prove that in the Bianchi class A, up to an
arbitrary homothetic factor
$\Ga>0$, there exist four, and only four, one-parameter families
of complete Einstein-Weyl metrics with a non-exact Weyl form. More
precisely, given a positive conformal scalar curvature
$S^D \equiv 4\Ga l_2$
\footnote{\ Looking for complete metrics, we explicitely checked a
theorem given by Pedersen and Swann
\cite{{PDSRheine},{Tod92},{Gauduchon}}, asserting that a compact
non-exact four-dimensional Einstein-Weyl space should have a
strictly positive conformal scalar curvature.}, the parameters in
equation (\ref{3-47}) are fixed and in particular the metrics are
Bianchi $IX$ ones. The absence of any complete metrics in Bianchi
$VII_0$ and $VIII$ subclasses agrees with
 refs.
\cite{{Madsen-a},{mpps}} that show that the compactness of the
metric requires a compact symmetry group. The four families are
respectively nut-nut, nut-bolt, bolt-nut or bolt-bolt.
\nl\noindent  The paper is organised as follows :  in the next
Section, we study the regularity for Bianchi $VIII$ or $IX$
metrics and discuss their completeness. For Bianchi $VII_0$
candidates, we show in Section 3 that they cannot be regular and
finally, after a comparison with previous results
\cite{{PDSRheine},{Madsen-a},{mpps}}, conclude with some comments
on the topological properties of our compact solutions and the
underlying manifolds in Section 4.

\section{Bianchi \boldmath $VIII$ and $IX$ complete metrics ?}
Let us introduce the new ''proper time "
\beq\label{n3} t = k -n^{33}x \,. \eeq The distance becomes :
\beqa\label{n31} ds^2 & = & \frac{2}{\Ga}\left
[\frac{t}{\tilde{\Om}^2(t)[1 + (t-k)^2]^2}(dt)^2
 + \frac{t}{[1 + (t-k)^2]}[(\si^1)^2 + (\si^2)^2] +
\frac{\tilde{\Om}^2(t)}{t}(\si^3)^2 \right] \ ,\nnb \\
\tilde{\Om}^2(t)  & = & n^{33} + (n^{33}l_1 - l_2
\arctan(t-k))[t^2 - (1+k^2)]
 - l_2(t+k)
\eeqa In our local approach, the ``time'' interval results from
the positivity of the distance (\ref{n31}) : this leads to the two
conditions
$$t > 0\ \,,\
\tilde{\Om}^2(t) > 0\,.$$

\noindent As
\beq\label{n32}
\frac{d\tilde{\Om}^2(t)}{dt} = 2tg(t)\ ,\ \ g(t) = n^{33}l_1 -
l_2[\arctan(t-k) +\frac{t-k}{1+(t-k)^2}]\ ,\ \frac{dg(t)}{dt} =
\frac{-2l_2}{[1+(t-k)^2]^2}
\eeq
 the discussion separates into three subcases according to the
behaviour of
$\tilde{\Om}^2(t) $ when $t$ varies. Let $$\de = g(+\infty) =
n^{33}l_1 -
\frac{\pi}{2}l_2 \ {\rm and}\ \ \mu = g(0) =  n^{33}l_1 +
l_2(\arctan{k} +\frac{k}{1+k^2})\,:$$
\begin{itemize}
\item a) $\de \ge 0\ \ {\rm and}\ \ \mu \ge 0$ : $\tilde{\Om}^2(t)
$ monotonicaly increases with $t$ .

\item b) $\de \le 0\ \ {\rm and}\ \ \mu \le 0$ : $\tilde{\Om}^2(t)
$ monotonicaly decreases with $t$ .

\item c) $\de\mu < 0\,$ : $\tilde{\Om}^2(t) $ has a single
extremum when $t$ varies (obtained when g(t) vanishes). The\
extremum is a maximum if $l_2\ {\rm and}\
\mu > 0\,$(case $ c^+$) and a minimum if $l_2\ {\rm and}\ \mu <
0\,$(case $ c^-$).
\end{itemize}  The behaviour of the distance at
$ \infty$  is readily seen to be singular if
$\de \neq 0\,.$ Indeed,
\beq\label{n1}
 ds^2 \sim  \frac{2}{\Ga}\left[\frac{(dt)^2}{\de t^5} +
\frac{1}{t}[(\si^1)^2 + (\si^2)^2] +\de t(\si^3)^2\right]\,,\ \ t
\rightarrow +\infty\,,
\eeq
 and the change of variable $\rho = (t)^{-3/2}$ leaves {\bf a non
removable singularity} at $\rho = 0\,.$

Consider now the special case when $\de$ vanishes : the function
$\tilde{\Om}^2(t)$ goes to $n^{33}$ when $t
\rightarrow +\infty\,.$ So,
$ n^{33}$ should be +1, and the distance behaves as
\beq\label{n11}  ds^2 \sim
\frac{2}{\Ga}\left[\frac{(dt)^2}{t^3} +\frac{1}{t} [(\si^1)^2 +
(\si^2)^2 + (\si^3)^2\right]\,\ \ t \rightarrow +\infty\,,
\eeq  which gives a {\it nut} \cite{GH} after the change $\rho =
t^{-1/2}$:
$$ds^2 \sim \frac{8}{\Ga}\left[(d\rho)^2 +
\frac{\rho^2}{4}[(\si^1)^2 + (\si^2)^2 + (\si^3)^2]\right]\ \ ,\
\rho \rightarrow 0\,.$$ The singularity is removable if one choses
cartesian coordinates rather than polar ones. Note that in other
gauges, the metric is still regular, but, for example in the
``K\"ahler gauge", a boundary appears at infinity as the metric is
asymptotically euclidean:
$$d\tilde{s}^2 = \frac{\Ga}{2}[1+(t-k)^2]ds^2
\sim \frac{(dt)^2}{t} + t[(\si^1)^2 + (\si^2)^2 + (\si^3)^2] =
4(d\sqrt{t})^2 + (\sqrt{t})^2[(\si^1)^2 + (\si^2)^2 + (\si^3)^2]
\,,\ t\ \rightarrow \infty\ .$$   To sum up, we have
\nl\noindent{\bf Lemma 1 :} {\it if the proper time interval
extends to
$+\infty$, the metric can be regular only in the Bianchi $IX$ case
with $l_1 =
\frac{\pi}{2}l_2\,.$}
$$   $$ Consider now the behaviour of the distance near $t = 0$
supposed to be the lowest possible value of the proper time
compatible with a positive metric. If $\tilde{\Om}^2(0) > 0$, the
distance write - after the change of variable
$\rho = (t)^{3/2}$ :
\beq\label{n2}  ds^2 \sim
\frac{2}{\Ga}\left[\frac{4 (d\rho)^2}{9\tilde{\Om}^2(0)[1+k^2]^2}
+\frac{\rho^{2/3}}{[1+k^2]} [(\si^1)^2 + (\si^2)^2] +
\frac{\tilde{\Om}^2(0)}{\rho^{2/3}}(\si^3)^2
\right]\,,\ \rho \rightarrow 0\,,
\eeq  and is singular. Then we are left with the case
$\tilde{\Om}^2(0) = 0$, where one finds :
$$\frac{d\tilde{\Om}^2}{dt}(0) = 0\,,\ \ \
\frac{d^2\tilde{\Om}^2}{dt^2}(0) =
\frac{2n^{33}}{1+k^2}\,,$$ and here again the positivity requires
$n^{33} = +1\,.$ Then one can change the variable $t$ into $\rho$
given by
$$\rho =\sqrt{t}$$ and get the following {\it nut} behaviour for
the distance near $t = 0\ :$
$$ds^2 \sim \frac{8}{\Ga[1+k^2]}\left[(d\rho)^2 +
\frac{\rho^2}{4}[(\si^1)^2 + (\si^2)^2 + (\si^3)^2]\right]\,,\
\rho \rightarrow 0\,. $$

$$   $$

\noindent At last, singularities in the distance may occur at
zeroes of
$\Om^2(x)$ and, the range of variation of the proper time being
limited by such zeroes, one has to discuss the nature of the
singularity at these values, still using the work of Gibbons and
Hawking on {\it nuts} and {\it bolts} \cite{GH} to check the
completeness of the metric at both ends.

If $\tilde{\Om}^2(\al) = 0$ with
$\frac{d\tilde{\Om}^2}{dt}(\al) = 0$ and $\al
\neq 0$, positivity enforces $\al$ to be a minimum ; then,
$\tilde{\Om}^2(0) > 0$ which, as explained above,
 gives a singularity. So, consider the situation with
$\frac{d\tilde{\Om}^2}{dt}(\al) \neq 0$ and change the variable
$t$ to
$\rho$ according to :
$$ t = \al + \rho^2
\frac{d\tilde{\Om}^2}{dt}(\al)\ ;$$  using
$\tilde{\Om}^2(t)
\sim \rho^2[\frac{d\tilde{\Om}^2}{dt}(\al)]^2\,,$ the distance
behaves when
$\rho
\rightarrow 0\ $ as :
\beqa\label{n21}    ds^2 \sim
\frac{8\al}{\Ga[1+(\al - k)^2]^2} [(d\rho)^2 & + &
\frac{\rho^2}{4}
\left(\frac{1+(\al -
k)^2}{\al}\frac{d\tilde{\Om}^2}{dt}(\al)\right)^2(\si^3)^2 +
\nnb \\
 & + & \frac{1+(\al - k)^2}{4}[(\si^1)^2 + (\si^2)^2]]\ ;
\eeqa  and one has a {\it bolt of twist $ p$} iff. :
$$\frac{1+(\al - k)^2}{\al}\frac{d\tilde{\Om}^2}{dt}(\al) = \pm
p\,,\ \ p = 1,\ 2...$$ Indeed, in such a case, restricting the
range in the angle $\psi$ coresponding to
$\si^3 = d\psi + \cos\te d\phi\,,$ e.t.c... to the interval $[0,\
4\pi/p ]$, there is no longer a polar singularity in the distance.
This condition corresponds to :
\beq\label{n22} g(\al) = \pm \frac{p}{2[1+(\al - k)^2]}\ \ ,\ \ p
=1,\ 2,...\eeq
$$  $$ We have now all the building blocks needed in our
discussion on the regularity of our Einstein-Weyl metrics.

$$    $$

First, the absence of a non removable singularity at $t = 0$
supposes that :
\beq\label{n24}
\tilde{\Om}^2(0) \le 0 \Leftrightarrow n^{33}l_1 + l_2(\arctan k +
\frac{k}{1+k^2}) \ge \frac{n^{33}}{1+k^2}\,.
\eeq We now study successively the cases a), b),
$c^+$) and $c^-$).
\begin{itemize}
\item {\bf case a) : \boldmath $\tilde{\Om}^2(t)$ monotonicaly
increasing} : \nl\noindent the range of $t$ is
$[\al,\ +\infty[\ \ ;\ \al \ge 0\,,\
\tilde{\Om}^2(\al) = 0\,. $
\nl\noindent Lemma 1 enforces :
$ n^{33} = +1,\ l_1 = \frac{\pi}{2}l_2\ ;$ then we have a {\bf
nut} at $\infty\,.$ Moreover, thanks to (\ref{n24}), the relation :
$$ l_1 + l_2(\arctan k +
\frac{k}{1+k^2}) \equiv l_2(\frac{\pi}{2}+
\arctan k +
\frac{k}{1+k^2}) \ge \frac{1}{1+k^2}$$ implies
$l_2 > 0$, in agreement with the aforementioned theorem
\cite{PDSRheine}.
\begin{itemize}
\item {\bf \boldmath $\al = 0\ :$} we have a {\bf nut-nut} family
of metrics (depending on a positive parameter
$l_2$), whose distance is  given by the expression (\ref{3-47})
with :
\beqa\label{n34}  n^{33}  =  +1\, & , & \ l_1 =
\frac{\pi}{2}l_2\ ; \\ k\ {\rm given\ by\ the\ unique\ solution\
of}\ & :&\frac{\pi}{2}+ \arctan k + \frac{l_2 k -1}{l_2(1+k^2)} =
0\ .\nnb
\eeqa
\noindent In particular $k > -l_2\,.$

\item {\bf \boldmath $\al > 0\ :$} a complete metric needs $\al$
being a nut or a bolt of twist 1. Here, as $\al > 0\,,\
\frac{d\tilde{\Om}^2}{dt}(\al) > 0$ and only a bolt is possible :
we have a {\bf bolt-nut} family of metrics (depending on a
positive parameter $l_2$). The bolt condition (\ref{n22}) and the
vanishing of
$\tilde{\Om}^2(\al)$ uniquely determine the values of $\al\ {\rm
and}\ k$ and
 we have the following values for the parameters in the distance
(\ref{3-47}) :
\beqa\label{n48} n^{33} & = & +1\,,\ l_1 =
\frac{\pi}{2}l_2\,,\ \ k = \frac{1 + 4l_2 x_1 +  3x_1^2}{2(2l_2 +
x_1)}\ ,\ \ \al = k - x_1\ ,\\ x_1 & > & -2l_2 \ {\rm is\ the\
unique\ solution\ of\ }\ :\
\frac{\pi}{2} + \arctan x_1 + \frac{2l_2 x_1 - 1}{2l_2(1+ x_1^2)}
= 0\,,\nnb
\eeqa In particular, the position of the {\it bolt} is given by
$$\al =\frac{1+x_1^2}{2(2l_2 + x_1)}\ \ge
\frac{1}{2l_2 + \sqrt{4l_2^2 +1}}\,.$$

\end{itemize}

\item {\bf case b) : \boldmath $\tilde{\Om}^2(t)$ monotonicaly
decreasing} : \nl\noindent the distance cannot be positive and
this situation is excluded.

\item  {\bf case  \boldmath $c^+$) : \boldmath
$\tilde{\Om}^2(t)$ increases to a maximum value} reached for
$t = \be\,,\ g(\be) = 0\,,\ \tilde{\Om}^2(\be) > 0$ then decreases
and vanishes at
$\al'\,,\
\tilde{\Om}^2(\al') = 0\,;$ the range of $t$ is
$[\al,\ \al']\,.$ Moreover, a positive maximum requires $l_2$ and
$n^{33} > 0\,,$ i.e. $n^{33} = +1\,.$
\begin{itemize}

\item {\bf \boldmath $\al =0\,:$} we have a {\bf nut-bolt} of
twist 1 (replace $\pm p$ in equ.(\ref{n22}) by - 1) family of
metrics (depending on a positive parameter
$l_2$). The previous discussion enforces :
$$n^{33} = +1\ ,\ \mu = n^{33}l_1 + l_2(\arctan{k}
+\frac{k}{1+k^2}) =
\frac{1}{1+k^2}\ ,\ \de = l_1 - \frac{\pi}{2}l_2\ <\ 0\,.$$ The
existence of
$k$ then requires $ l_1 +\frac{\pi}{2}l_2\ >\ 0\ .$ Finally,
adding the bolt condition (\ref{n22}) and the vanishing of
$\tilde{\Om}^2(\al')$ , $\al'\ {\rm and}\ k$ are uniquely
\footnote{A useful result in the unicity discussion is the
expression of the derivative with respect to $x_2$ of the left
hand side of the equation in the third line of (\ref{n58}) : it
equals $\frac{19(2l_2 - x_2)}{l_2[9 + (x_2 + 4l_2)^2]^2}\,.$}
determined and
 we have the following values for the parameters in the distance
(\ref{3-47}) :
\beqa\label{n58}
 n^{33} = +1\ ,\ l_1 & = & -l_2[\arctan{x_2} +\frac{2l_2x_2
+1}{2l_2(1+x_2^2)}]\ ,\ k =
\frac{3 + 4l_2 x_2 + x_2^2}{2(2l_2 - x_2)}\ ,\ \
\al' = k - x_2\nnb \\ x_2 & < & 2l_2 \ {\rm is\ the\ unique\
solution\ of\ :}\nnb \\
\arctan x_2 & - &\arctan[\frac{3 + 4l_2 x_2 + x_2^2}{2(2l_2 -
x_2)}] +
\frac{3 + 2l_2 x_2 + 8l_2^2}{6l_2[1 + (\frac{x_2 + 4l_2}{3})^2]} =
0\,.
\eeqa   In particular, the position of the {\it bolt} is given by :
$$\al' =\frac{3(1+x_2^2)}{2(2l_2 - x_2)}\ \ge
\frac{3}{2l_2 + \sqrt{4l_2^2 +1}}\,.$$ Notice that with $x_2
\equiv -(4l_2 + 3 x_1)$ the equations (\ref{n58}) become :
\beqa\label{n59}\frac{\pi}{2} +
\arctan{x_1} +
\frac{2l_2 x_1 - 1}{2l_2(1+ x_1^2)} & = & 0\ \ ,\\
 k  =
\frac{1+ 4 l_2 x_1 +3x_1^2}{2(2l_2 + x_1)}\ \ ,\ \ l_1 & = & - l_2
[\arctan{k} + \frac{k l_2 - 1}{l_2(1 + k^2)}] \nnb
\eeqa with $x_1 > -2l_2\,.$ One recognises equ.(\ref{n48}), but
for the value of
$l_1\,.$ So, not too surprisingly, the metrics with a bolt at one
end and a nut at the other  end depend on a single transcendental
equation (\ref{n48}) which characterises the presence of a bolt of
twist 1.

\item {\bf \boldmath $\al>0\,:$} we have a {\bf bolt(p)-bolt(p)}
family of metrics (depending on a positive parameter $l_2$). The
previous discussion enforces :
$$n^{33} = +1\ ,\ \mu = n^{33}l_1 + l_2(\arctan{k}
+\frac{k}{1+k^2}) >
\frac{1}{1+k^2}\ ,\ \de = l_1 - \frac{\pi}{2}l_2\ <\ 0\,.$$ We
have two bolt conditions (\ref{n22}) : one with a $+p$ at $t =
\al\,,$ the other with a $-p$ at
$t = \al'\,.$ The vanishing of
$\tilde{\Om}^2(\al)$ and
$\tilde{\Om}^2(\al')$ , will determine $p\,,\ l_1\,,\ \al\,,\
\al'\ {\rm and}\ k$. As in the previous situation, we obtain :
\beqa\label{n49} l_1 & = & - l_2[\arctan{x_1} +\frac{2l_2x_1 -
p}{2l_2(1+x_1^2)}]\ ,\ k = x_1 -
\frac{(p-2)(1+x_1^2)}{2(px_1 + 2l_2)}\ ,\ \al = k -x_1 > 0\ ,\\
l_1 & = & - l_2[\arctan{x_2} +\frac{2l_2x_2+ p}{2l_2(1+x_2^2)}]\
,Ê\ k = x_2 -\frac{(2+p)(1+x_2^2)}{2(px_2 -2l_2)}\ ,\ \al' = k
-x_2 > \al > 0\ .\nnb
\eeqa
\begin{itemize}

\item The special value
$p=2$ enforces
$x_1 = -l_2$ then
$l_1 = 1+l_2\arctan (l_2)$. However, in that situation, the
equation that determines
$x_2$ has no solution.
\item The difference of the two lines of equations (\ref{n49})
leads to the vanishing of two functions :
\beqa\label{n50} f_1(x_1,x_2) & \equiv & -\arctan{x_1} +
\arctan{x_2} -\frac{2l_2x_1 -p}{2l_2(1+x_1^2)} + \frac{2l_2x_2
+p}{2l_2(1+x_2^2)} = 0\ ,
\nnb
\\  f_2(x_1,x_2) & \equiv & x_1 - x_2 -
\frac{(p-2)(1+x_1^2)}{2(px_1 + 2l_2)}
+\frac{(2+p)(1+x_2^2)}{2(px_2 -2l_2)} = 0\ ,
\eeqa For $p>2$, $\al' > \al > 0$ requires
$x_1< - 2l_2/p$ and
$x_2< 2l_2/p\,.$ In this domain, the two partial derivatives of
$f_1(x_1,x_2)$ are positive ; on the other hand
$$f_1(x_1 = x_2,\ x_2) =
\frac{p}{l_2(1+x_2^2)} >0\ :$$
 as a consequence,
$f_1(x_1 > x_2,\ x_2) > f_1(x_1 = x_2,\ x_2)$ cannot vanish. Then
we have

\vspace{0.3cm}

\noindent{\bf Lemma 2 :} {\it Regular bolt-bolt Einstein-Weyl
Bianchi $IX$ metrics, non conformally Einstein, exist only with a
twist
$p = 1\,.$}

\vspace{0.3cm}

\item For $p=1$, equation $f_2(x_1,x_2) = 0$ gives either $x_1 =
-\frac{4l_2 +x_2}{3}$ or $x_1 =
\frac{2l_2 x_2 +1}{2l_2 - x_2}\,.$
\nl\noindent With regards to
$f_1(x_1,x_2)
$  , when computing its total derivative with respect to
$x_2$ with either solution for $x_1$, one readily finds that
$f_1(\frac{2l_2 x_2+ 1}{2l_2 - x_2},\ x_2) $ is the positive
constant :
$\left[1/(2l_2) -
\arctan(1/(2l_2))\right]$, which excludes that
solution.\nl\noindent On the contrary,
$f_1(-\frac{4l_2 + x_2}{3},\ x_2)$ being monotonically increasing
from $-\pi$ when
$x_2 \rightarrow -\infty$ to a positive value when $x_2= 2l_2\,,$
one obtains {\bf one and only one} solution.

\vspace{0.3cm}

\noindent To sum up, we have the following values for the
parameters in the distance (\ref{3-47}) :
\beqa\label{n52}
 n^{33} = +1\ ,\ l_1 & = & -l_2[\arctan{x_2} +\frac{2l_2x_2
+1}{2l_2(1+x_2^2)}]\ ,
\ k =  \frac{3 + 4l_2 x_2 + x_2^2}{2(2l_2 - x_2)}\ ,\nnb \\ {\rm
where\ }\ x_2 & < & 2l_2
\ {\rm is\ the\ unique\ solution\ of\ :}\ f_1(-\frac{4l_2
+x_2}{3},\ x_2) = 0\,.
\eeqa  The positions of the {\it bolts} are given by
$$\al = \frac{9+ (4l_2 + x_2)^2}{6(2l_2 - x_2)}
\ge \frac{1}{2l_2 +
\sqrt{4l_2^2 +1}}\\ ,\ \al' =\frac{3(1+x_2^2)}{2(2l_2 - x_2)} \ge
\frac{3}{2l_2 +
\sqrt{4l_2^2 +1}}\\,.$$

\end{itemize}
\end{itemize}
\item {\bf case \boldmath $c^-$) : \boldmath
$\tilde{\Om}^2(t)$ decreases to a minimum value} reached for
$t = \be\,,\ g(\be) = 0\,,$ then increases and vanishes at
$\al'\,,\
\tilde{\Om}^2(\al') = 0\,;$ the range of $t$ then goes to
$+\infty$ which, requiring $\de = 0$, contradicts $\de\mu < 0\,.$
\end{itemize}

\noindent {\bf To summarize}, in our Gauduchon's gauge we found
four, and only four, families of regular
 Einstein-Weyl Bianchi
$IX$ metrics;  according to the classification of Gibbons and
Hawking, they are complete and live on a compact manifold without
boundary (see also in the concluding section the expression of the
volume of the manifold) and have a constant positive conformal
scalar curvature in agreement with the quoted theorem.

\section{ Bianchi \boldmath$VII_0$ regular metrics ?} The metric
write :
\beqa\label{n41}     ds^2 & = &
\frac{2}{\Ga}\left [\frac{k}{\Om^2(x)[1 + x^2]^2}(dx)^2
 + \frac{k}{[1 + x^2]}[(\si^1)^2 + (\si^2)^2] +
\frac{\Om^2(x)}{ k}(\si^3)^2 \right] \ ,\nnb \\
\Om^2(x)  & = & -2k[x(l_1 + l_2\arctan x) + l_2]\,.
\eeqa The positivity of the distance (\ref{n41}) enforces the two
conditions
$$k > 0\ \,,\
\Om^2(x) > 0\,.$$ As
\beq\label{n62}
\frac{d\Om^2(x)}{dx} = - 2kh(x)\ ,\ \ h(x) = l_1 + l_2[\arctan x
+\frac{x}{1+x^2}]\ ,\
\frac{dh(x)}{dx} = \frac{2l_2}{[1+x^2]^2}\ ,
\eeq here also the discussion separates into three subcases
according to the behaviour of
$\Om^2(x) $ when $x$ varies. Let $\de^{\pm} = h(\pm\infty) =
  l_1 \pm
\frac{\pi}{2}l_2\ :$
\begin{itemize}
\item a) $
\de^{\pm} \le 0$ : $\Om^2(x) $ increases with
$x\,.$
\nl\noindent The proper time $x$ belongs to the interval
$[\al\,,\ +\infty[\,,$ where
$\Om^2(\al) = 0\,.$
\item b) $\de^{\pm} \ge 0$ : $\Om^2(x) $ decreases with $x\,.$
\nl\noindent The proper time
$x$ belongs to the interval
$]-\infty\,,\ \al]\,.$ where
$\Om^2(\al) = 0\,.$

\item c) $\de^+\de^- < 0$ : $\Om^2(x) $ has a single extremum when
$x$ varies (obtained when
$h(x)$ vanishes).

 The extremum is a maximum if $ l_2 > 0\,$(case $ c^+$), a minimum
if $ l_2 < 0\,$(case $ c^-$).
\nl\noindent In the first case the maximum is readily seen to lead
to $\Om^2 < 0\,;$ one remains with the sole case $c^-$, where the
range of the proper time covers the whole $x$ axis
$]-\infty\,,\ +\infty[\,.$
\end{itemize}

\noindent The behaviour of the distance at
$\pm \infty$  is readily seen to be singular if
$\de^{\pm} \neq 0\,.$ Indeed,
\beq\label{n61}
 ds^2 \sim
\frac{2}{\Ga}\left[\frac{(dx)^2}{(-2\de^{\pm}) x^5} +
\frac{k}{x^2}[(\si^1)^2 + (\si^2)^2] -2\de^{\pm}
x(\si^3)^2\right]\,,\ \ x
\rightarrow
\pm\infty\,,
\eeq
 and the change of variable $\rho = (\pm x)^{-3/2}$ leaves a non
removable singularity at
$\rho = 0\,.$

To sum up, we have \nl\noindent{\bf Lemma 3 :} {\it in the Bianchi
$VII_0$ case, if the proper time interval extends to
$+\infty$ (resp. $-\infty$), the metric can be regular only if
$l_1 +
\frac{\pi}{2}l_2 = 0$ (resp. $l_1 -
\frac{\pi}{2}l_2 = 0$)}

Then the function
$\Om^2(x) $ simplifies to :
$$\Om^2(x) = -2kl_2[1 \mp x (\frac{\pi}{2} \mp
\arctan x)] \sim -2k l_2[1 - x\arctan(\frac{1}{x})]\ \ {\rm when}\
x
\rightarrow \pm\infty\,.$$ We now study successively the cases a),
b) and $c^-$.
\begin{itemize}
\item {\bf case a) :  \boldmath $\Om^2(x)$ monotonicaly
increasing} : only the case
$\de^+ = 0$ with $\de^- < 0\,,\ i.e.\ l_2 >0$  has to be
considered, then the function
$\Om^2(x)$ is always
$< 0\,,$ which is excluded.
\item {\bf case b) : \boldmath $\Om^2(x)$ monotonicaly decreasing}
: only the case
$\de^- = 0$ with $\de^+ > 0\,,\ i.e.\ l_2 >0$ has to be
considered, but here again the function
$\Om^2(x)$ is always $< 0\,,$ which is excluded.

\item {\bf case \boldmath $c^-$) : \boldmath
$\tilde{\Om}^2(t)$ decreases to a minimum value} then increases.
One of the two ends of the $x$ axis remains singular.
\end{itemize}

\noindent Then there is no regular Einstein-Weyl metric in the
Bianchi $VII_0$ class.

\section {Concluding remarks} In this paper, we have presented an
analysis of complete Einstein-Weyl structures ($g\,,\
\ga$) corresponding to Class A non-conformally Einstein Bianchi
metrics in a Gauduchon's gauge. Thanks to our previous results,
they are conformally K\"ahler. Let us recall that the extra
$U(1)$ enlarging the Bianchi isometry group corresponds to the Killing vector
dual to the Weyl form
$\ga\,,$ and is responsible for the conformally K\"ahler property.

Firstly, we have found that there are no regular metric in
the Bianchi
$VII_0$ and $VIII$ subclasses ; this is not so surprising as in
those cases, the underlying Lie group is non-compact :
$U(1)\times E(3)$(group of displacements in 3 dimensions) or
$U(1)\times SU(1,1)$ respectively. Then, according to
\cite{{Madsen-a},{mpps}} there cannot exist compact Einstein-Weyl
solutions. On the contrary, Bianchi
$IX$ Einstein-Weyl metrics have
$U(1)\times SU(2)$ as isometry group which allows for a compact
solution.

Secondly, we have shown how in each of the four
families obtained, the completeness requirement determines all the
parameters, but the homethetic one
$\Ga >0$, as soon as the (positive) conformal scalar curvature is
fixed.

Let us now comment on their topological properties.
\begin{itemize}

\item Notice that in a Gauduchon's gauge, the Ricci tensor
of the Levi Civita connection is positive definite as :
$$R_{ij} = g_{ij}S^D /4 +(g_{ij}\mid\mid\ga\mid\mid^2 -
\ga_i\ga_j)/2\,,$$ and the scalar curvature, related to the
conformal scalar one through :
$$R = S^D +\frac{3}{2}\mid\mid\ga\mid\mid^2\
\Rightarrow \ R = 4\Ga[ l_2 +
\frac{\tilde{3\Om}^2(t)}{4t}]\ ,$$ is strictly positive on the
manifold. As a consequence
$\pi_1$ is finite and
$b_1$ vanishes \cite{PDSRheine}.

\item The volume element writes :
\beq\label{c1} d{\cal V} =
\frac{4t}{\Ga^2[1+(t-k)^2]^2}dt\sin{\te}d\te d\Phi d\Psi\
\Rightarrow {\cal V} =  (\frac{4\pi}{\Ga})^2
\frac{V}{l_2}   \,,
\eeq with $V =\frac{2l_2 + 2k}{1+k^2}$ (solution (\ref{n34})), $V
=\frac{2l_2 +k}{1+x^2}$ (solutions (\ref{n48},\ref{n58})) and $V =
\frac{2l_2 + k}{1+x^2} - \frac{2l_2 - k}{1+x'^2}$ (solution
(\ref{n52})); this confirms the compact character of our metrics.
But, if one goes to the K\"ahler gauge through the conformal
transformation :
$$ \tilde{g} =
\frac{\Ga[1+(k-t)^2]}{2}g\ ,\ \ \tilde{\ga} =
\frac{\tilde{2\Om}^2(t)}{2t}\si^3 +
\frac{2(k-t)dt}{1+(k-t)^2}\,,$$ the volume element becomes :
$$ d\tilde{{\cal V}} = t dt\sin{\te}d\te d\phi d\Psi\ ,$$ and only
the two families with a finite range for the proper time stay
compact. As explained in Section 2 (after equation (\ref{n11})),
for the first two families, an asymptotically euclidean boundary
appears at $\infty$.
\item We have also computed the Euler number
$\chi_{\sc e}$ and the signature $\tau$ for the compact metrics
(\ref{n34},\ref{n48},\ref{n58},\ref{n52}) and found respectively :
$\chi_{\sc e}$ = 2, 3, 3 or 4, and $\tau$ = 0, -1, +1 or 0, in
agreement with
\cite{EH} as a nut contributes 1 to the Euler number whereas a
bolt contributes 2. For a compact four-dimensional manifold, $b_0
= b_4 =1$ and $b_1 = b_3$; moreover, here $b_1$ vanishes, then one
gets $$\chi_{\sc e} = 2 + b_2^+ + b_2^-\ ,\ \ \tau = b_2^+ -
b_2^-\,,$$ where $b_2^{\pm}$ is the number of self(anti-self)dual
harmonic 2-forms on the manifold. The candidates harmonic 2-forms
are :
\beqa\label{c3}
\om^+ & = & (dt\wedge \si^3 + t\si^1\wedge \si^2) =
\frac{\Ga[1+(t-k)^2]}{2}[e_0\wedge e_3 + e_1
\wedge e_2]\ ,\\
\om^- & = & (dt\wedge
\si^3 - t\si^1\wedge \si^2)/(t^2) =
\frac{\Ga[1+(t-k)^2]}{2t^2}[e_0\wedge e_3 - e_1
\wedge e_2]\ .
\nnb
\eeqa Analysing the possible singularities of these 2-forms at the
end points of the proper time interval, one easily sees that :
\begin{itemize}
\item our nut-nut family (\ref{n34}) has
$ b_2^+ = b_2^- = 0$ as the two 2-forms (\ref{c3}) are singular at
$t=0$ or at
$t
\rightarrow +
\infty$, then $\chi_{\sc e}$ = 2 and $\tau$ = 0 ;
\item in the same way, the bolt-nut family (\ref{n48}) has $ b_2^+
= 0$ and $b_2^- = 1$ ,  then
$\chi_{\sc e}$ = 3 and $\tau$ = -1 whereas the nut-bolt family
(\ref{n58}) has $ b_2^- = 0$ and
$b_2^+ = 1$ ,  then $\chi_{\sc e}$ = 3 and $\tau$ = +1 (as
remarked previously, these two families are similar but the
orientation of the manifolds are opposite);
\item finally,  the bolt-bolt family (\ref{n52}) has $ b_2^+ =
b_2^- = 1$, then $\chi_{\sc e}$ = 4 and $\tau$ = 0.
\end{itemize}
 Q.E.D.
\end{itemize}

\noindent Let us finally compare our results with thoses of
Pedersen and Swann
\cite{PDSRheine} and the systematic analysis of compact
Einstein-Weyl structures with a large symmetry group done by
Madsen in his PHD dissertation (\cite{Madsen-a}-especially Section
8), completed in \cite{mpps}(especially section 7) \footnote{\
When a first version of this work was completed, I was not aware
of this publication, which explains some of the new comments. I
thank the referees who drew my attention to this mathematical
publication.}. The ``time parameter" in these analyses is an
angle $\varphi\,.$

Of course, our results are not new when compared to the global
mathematical approach of \cite{{Madsen-a},{mpps}}, but we think
our language is more relevant for physicists and, moreover, our
parametrisation being simpler, our discussion is able to correct
some important mistakes.

\begin{itemize}

\item As the analysis of Gibbons and Hawking in the language of
bolts and nuts applies only to {\bf orientable manifolds}, it is
not surprising that, contrary to \cite{{Madsen-a},{mpps}}, we do
not find as solutions, metrics on non-orientable manifolds such as
$RP^4$ or
$RP^4 \# CP^2\,.$

\item Our {\bf nut-nut} family (\ref{n34}) corresponds to the
solution where the manifold is
$S^4\,,$ the ``time parameter" $\varphi \in [\varphi_0,\ \pi -
\varphi_0]$ and {\it special orbits at each end being a point}
(subsection 8-20 of
\cite{Madsen-a}, case 1, page 426 of \cite{mpps}).

\item Our {\bf nut-bolt} families  (\ref{n48},\ref{n58})
correspond to the solution where the manifold is
$CP^2$ and {\it special orbits are respectively a point and
$CP^1$} (subsection 8-22 of
\cite{Madsen-a}) : the (\ref{n48}) case corresponds to a ``time
parameter"
$\varphi \in [\varphi_0,\ \Phi]$, special orbits being a point at
$\varphi_0 $ and  $CP^1$ at $\Phi$ (in the notations of Madsen) ;
the (\ref{n58}) case corresponds to a ``time parameter" $\varphi
\in [\Phi,\ \pi -\varphi_0]$, special orbits being a point at $\pi
- \varphi_0 $ and  $CP^1$ at
$\Phi$ . In \cite{mpps}, an isolated solution is also given (case
3a, page 427) ; however, this is wrong : an easy calculation
checks that, contrary to their assertion, it does belong to the
previous one-parameter family.

\item Our {\bf bolt-bolt} family (\ref{n52}) corresponds to the
solution where the ``time parameter" $\varphi \in [\Phi_1,\
\Phi_2]$, and {\it special orbits are both $CP^1$} (subsection
8-24 of
\cite{Madsen-a}), but we have been able to prove that no
bolt(p)-bolt(p) exists with
$p\ge 2$ (Lemma 2), a result which was only conjectured by Madsen.
Moreover, thanks to our parametrisation that disentangles the
parameters
$k,\ x_1\ $ and $x_2$ into a single transcendental equation for
only one unknown parameter, we also proved that the relation
$\Phi_1 + \Phi_2 = \pi\ ,(c.f.\ 3x_1 + x_2 = -4l_2)$ , also
conjectured in Madsen's thesis dissertation, is indeed the sole
solution. In \cite{mpps}, it is also proven, with some efforts,
that the conjectured relation $\Phi_1 + \Phi_2 = \pi\ $ holds. The
authors claimed too that, apart from $p=1\,,$ solutions exist
for any $p \ge 4\,;$ however in their proof (pages 429-430), they
forget about a constraint on their parameter $\chi$ which has to
be at least equal to their parameter $D$; therefore, $\tan(\chi/2)$ ,
which they found must be greater than $\frac{2}{p}\tan(D/2)$,
should also be  greater than $\tan(D/2)$ which is a more
restrictive condition as soon as $p > 2\,.$ When this condition is
added, one finds that  no bolt(p)-bolt(p) exists with
$p\ge 2\,.$

As a final remark, note that the manifold is the Hirzebruch
surface $ CP^2
\#
\overline{CP^2}$, as in the family firstly exhibited by Pedersen
and Swann
\cite{PDSRheine}. Several other metrics are known to live on this
manifold (Page's Einstein metric
\cite{Page}, Calabi's extremal K\"ahler metric
\cite{Calabi}, Chave and Valent's quasi-Einstein K\"ahler metric
\cite{Valent96}).
\end{itemize}

Our results may be summarized in the \nl\noindent {\bf Theorem :} {\it In a
Gauduchon's gauge, non-exact Einstein-Weyl structures (g, $\ga$) corresponding
to class A (non-)diagonal Bianchi metrics on an orientable manifold are regular
only when [g] belongs to the Bianchi $IX$ subclass :
\nl i) the metrics are conformally K\"ahler, \nl ii) they have an arbitrary
constant positive conformal scalar curvature, \nl iii) they occur in four
one-parameter families characterised by their end-points being respectively
nut-nut, nut-bolt(1), bolt(1)-nut and bolt(1)-bolt(1), \nl iv) they live on the
compact manifolds $S^4$, $CP^2$ (with two possible orientations) and $CP^2\#
\overline{CP^2}$ respectively .}

\bibliographystyle{plain}

\end{document}